# Two major accretion epochs in M31 from two distinct populations of globular clusters

Dougal Mackey[1*], Geraint F. Lewis[2], Brendon J. Brewer[3], Annette M. N. Ferguson[4], Jovan Veljanoski[5], Avon P. Huxor[6], Michelle L. M. Collins[7], Patrick Côté[8], Rodrigo A. Ibata[9], Mike J. Irwin[10], Nicolas Martin[9,11], Alan W. McConnachie[8], Jorge Peñarrubia[4], Nial Tanvir[12], Zhen Wan[2]

[1]Research School of Astronomy & Astrophysics, Australian National University, Canberra, ACT 2611, Australia
[2]Sydney Institute for Astronomy, School of Physics, A28, The University of Sydney, NSW 2006, Australia
[3]Department of Statistics, The University of Auckland, Private Bag 92019, Auckland 1142, New Zealand
[4]Institute for Astronomy, University of Edinburgh, Royal Observatory, Blackford Hill, Edinburgh, EH9 3HJ, UK
[5]Kapteyn Astronomical Institute, University of Groningen, PO Box 800, NL-9700 AV Groningen, Netherlands
[6]HH Wills Physics Laboratory, University of Bristol, Tyndall Avenue, Bristol, BS8 1TL, UK
[7]Department of Physics, University of Surrey, Guildford, GU2 7XH, Surrey, UK
[8]NRC Herzberg Astronomy and Astrophysics, 5071 West Saanich Road, Victoria, B.C., V9E 2E7, Canada
[9]Université de Strasbourg, Observatoire astronomique de Strasbourg, UMR 7550, F-67000 Strasbourg, France
[10]Institute of Astronomy, University of Cambridge, Madingley Road, Cambridge, CB3 0HA, UK
[11]Max-Planck-Institut für Astronomie, Königstuhl 17, D-69117 Heidelberg, Germany
[12]Department of Physics & Astronomy, University of Leicester, University Road, Leicester LE1 7RH, UK

## Abstract

**Large galaxies grow through the accumulation of dwarf galaxies[1,2]. In principle it is possible to trace this growth history using the properties of a galaxy's stellar halo[3,4,5]. Previous investigations of the galaxy M31 (Andromeda) have shown that outside a radius of 25 kpc the population of halo globular clusters is rotating in alignment with the stellar disk[6,7], as are more centrally located clusters[8,9]. The M31 halo also contains coherent stellar substructures, along with a smoothly distributed stellar component[10,11,12]. Many of the globular clusters outside 25 kpc are associated with the most prominent substructures, while others are part of the smooth halo[13]. Here we report a new analysis of the kinematics of these globular clusters. We find that the two distinct populations are rotating with perpendicular orientations. The rotation axis for the population associated with the smooth halo is aligned with the rotation axis for the plane of dwarf galaxies[14] that encircles M31. We interpret these separate cluster populations as arising from two major accretion epochs, likely separated by billions of years. Stellar substructures from the first epoch are gone, but those from the more recent second epoch still remain.**

## Main

We conducted the Pan-Andromeda Archaeological survey (PAndAS) to obtain a multi-band panoramic view of the M31 stellar halo out to a radius of ∼ 150 kpc in projection[15,16]. As well as revealing extensive halo substructure from tidally disrupted dwarfs, PAndAS uncovered a substantial number of globular clusters at large galactocentric radii (Figure 1)[17,18,19]. The focus of the present study is the population of globular clusters in the outer halo of M31, which we define as the 92 objects known at projected radii larger than 25 kpc[13]. We have previously undertaken a series of spectroscopic observations to obtain line-of-sight velocities for the majority of this sample (77 of 92 clusters) with a typical measurement uncertainty per target of ∼ 10 km/s[6,7,13]. Treated as a single population, these objects exhibit net rotation about the centre of M31 with amplitude ∼ 85 km/s[6,7].

Our recent work[13] has allowed us to classify many of the clusters in our sample as being physically associated with underlying halo substructures (a subgroup hereafter referred to as *GC-Sub*), while others exhibit no association with substructure (hereafter *GC-Non*) and are plausibly part of the smooth halo. The remaining clusters are labelled "ambiguous". With this division in place, we revisit our understanding of the kinematics of

M31's outer globular cluster population. Our analysis is restricted to the region outside 25 kpc because at smaller radii our classification scheme is compromised by the high degree of overlapping substructure, plus extensive stellar material stripped from Andromeda's disk; moreover, the shorter dynamical times tend to more quickly erase any association between clusters and substructures.

We consider a family of three kinematic models to explain the observed cluster velocities, each consisting of an underlying rotational component and a superimposed velocity dispersion (see Extended Data Tables 1 and 2). For a given model, we examine two scenarios: first, a single global rotation plus dispersion to describe the entire globular cluster population, and then a solution comprising two distinct components corresponding to the GC-Sub and GC-Non subgroups and with the ambiguous objects treated in a probabilistic fashion. We employ Bayesian inference to compute the probabilities of the models given the data, and determine their veracity using Bayes factors. The results are summarised in Extended Data Table 3.

Examining these, it is immediately apparent that the outer halo globular cluster population of Andromeda exhibits significant global rotation, in close agreement with previous results[6,7]. However, it is also clear that, for each of the three models considered, the statistical evidence overwhelmingly favours the case where two rotational components are present over that with a single rotational component. That is, dividing the globular cluster population into two physically-motivated subgroups, each of which exhibits its own distinct rotational signature, yields a statistically significant improvement in explaining the observed data (as indicated by the >100-fold increase in Bayes factor in all three cases).

We turn to a brief description of the favoured model, $V_2$ in Extended Data Table 1, for which the inferred parameter values are listed in Extended Data Table 4 and the best-fit rotational kinematics are illustrated in Figure 2. In general, similar conclusions hold for all of the models considered in our analysis.

In $V_2$, the inferred amplitude of rotation for each of the two cluster subgroups is similar, and non-zero at $> 3.5\sigma$. In both cases it is comparable in magnitude ($\sim 100$ km/s) to the velocity dispersion. In the Cartesian coordinate system adopted for our fits (see Figure 1), the axis of rotation for GC-Non is oriented at $(-6.2 \pm 13.4)°$, whereas that for GC-Sub sits at $(80.0 \pm 10.0)°$. It is notable that these rotational axes are orthogonal to each other within the uncertainties. The projected angular momentum vectors for the two cluster subgroups are represented as arrows in Figure 2 (adopting the right-hand convention), and are oriented at $\sim 84°$ for GC-Non and $\sim 170°$ for GC-Sub in the standard astronomical coordinate system where angles are taken from North through East. For comparison, the rotation axis of Andromeda's stellar disk is oriented at $128°$ in the standard astronomical system, with a disk rotation amplitude of $\sim 250$ km/s and dispersion of $\sim 60$ km/s[20,21]. Inside 20 kpc, the population of metal-poor globular clusters rotates with a similar amplitude to that inferred for our cluster subgroups ($\sim 100$ km/s), but exhibits somewhat higher dispersion $\sim 150$ km/s[8,9]. The rotation axis for the inner metal-poor clusters is in the same sense as Andromeda's stellar disk, but plausibly misaligned by $\approx 20-30°$ in the direction of the GC-Non axis.

How are we to interpret the existence of rotational signatures with approximately perpendicular orientation amongst the outer globular cluster populations in Andromeda? The GC-Sub sample consists of clusters that are members of well-defined stellar streams or small dynamically cold subgroups in the halo. These clearly represent debris from one or more accretions that must have occurred relatively recently in order for the underlying structures to still be coherent[7,13,19]. It is well established that Andromeda has undergone at least one substantial late merger, accreting the fourth or fifth largest galaxy in the Local Group approximately one billion years ago and producing the Giant Stellar Stream in the process[22,23]. Presently-available models follow only the late stages of this system's infall, and suggest that at the point of destruction the progenitor's orbit was probably highly radial[23]. This makes it difficult to assess the degree to which the observed rotation of the GC-Sub sample could be due to this event.

Some recent analyses have advocated for a scenario where Andromeda experienced a much larger ($\sim 1:4$ mass ratio) merger event within the past $\sim 2-3$ billion years[24,25], based on more indirect lines of evidence. The high-amplitude rotational signal observed here for GC-Sub is certainly consistent with this overall picture, but we are presently unable to distinguish between the accretion of one large progenitor, or several smaller satellite systems with correlated angular momenta. One hypothesis arising from this scenario is that the compact elliptical satellite galaxy M32 may be the stripped core of the accreted progenitor[25]. If true, we predict that future astrometric measurements of M32's transverse motion will imply a projected angular momentum vector aligned with that for GC-Sub; however, we also note that no globular clusters are known to be bound to this satellite.

The GC-Non sample consists of clusters that exhibit no evidence for association with halo substructure. As such, they have previously been linked with the smooth halo component in Andromeda[13]. That these objects exhibit a substantial net angular momentum again suggests that they are plausibly due to either a single large merger or the accretion of several lesser progenitors with correlated orbits. However, this assembly epoch must have occurred billions of years earlier than the GC-Sub accretion(s) in order for phase mixing to erase coherent substructure over multiple orbital periods; some memory of the original large-scale angular momentum is retained during this process provided the shape of Andromeda's gravitational potential does not vary strongly with time.

It is particularly intriguing that the GC-Non sample rotates in the same sense, and is very closely aligned with, Andromeda's prominent plane of satellites[14,26]. The rotational axis of this plane, comprising nearly half of the dwarf galaxy population orbiting Andromeda, is oriented at an angle of $\approx -9.7°$ in the adopted Cartesian coordinates[26]. This is within a few degrees of the GC-Non axis, well within the uncertainty on both quantities. This close alignment is surprising because satellite planes are fragile, short-lived structures, destroyed by precession and interaction with dark matter halo substructure unless they are finely aligned within a spherical gravitational potential[27,28]. Assuming the properties of the GC-Non sample indeed represent an ancient accretion epoch, the progenitor systems cannot, therefore, have been members of the present-day plane. If the strong rotational alignment between the two structures is not coincidental, it might instead point to the larger-scale accretion of material onto Andromeda from a preferred direction in the local environment (e.g., along a dominant dark matter filament[29]) over cosmic history.

Can we deduce any properties of the destroyed GC-Sub and GC-Non progenitors from their now-assimilated globular cluster populations? Empirically, the overall mass of globular clusters in a galaxy is a fraction $\approx 2.9\times10^{-5}$ of the total mass of the system, that holds constant across a range of five orders of magnitude in host galaxy mass[30]. In our favoured model $V_2$ the total number of clusters in GC-Sub is 43, and in GC-Non is 34. These include probabilistically-assigned members of the ambiguous set. Assuming a median cluster mass $1.3\times10^5$ $M_\odot$[13,30], the implied progenitor masses are then $1.9\times10^{11}$ $M_\odot$ and $1.5\times10^{11}$ $M_\odot$ respectively. These are lower limits, as the accretion events almost certainly deposited additional clusters into Andromeda's halo at radii inside 25 kpc. As an individual accretion, the GC-Sub event would correspond to a $\sim$ 1:10 mass-ratio merger of a galaxy comparable to the Large Magellanic Cloud or the inferred Giant Stream progenitor[23]. The GC-Non event is similar, but having occurred much earlier in Andromeda's history would likely represent a rather more substantial relative merger.

The distinct rotational signatures we have uncovered for globular cluster subsystems in the outskirts of Andromeda offer a powerful insight into the assembly history of our nearest large neighbour. Our results provide empirical evidence that Andromeda has undergone at least two major epochs of satellite accretion – one relatively recent in cosmological terms, and one that likely occurred many billions of years ago. The apparent kinematic alignment between one or more long-destroyed ancient progenitor systems and the present-day plane of satellites may indicate long-term coherent accretion from the cosmic web. These represent important steps towards a detailed reconstruction of the main events that led to the Local Group of galaxies that we see today.

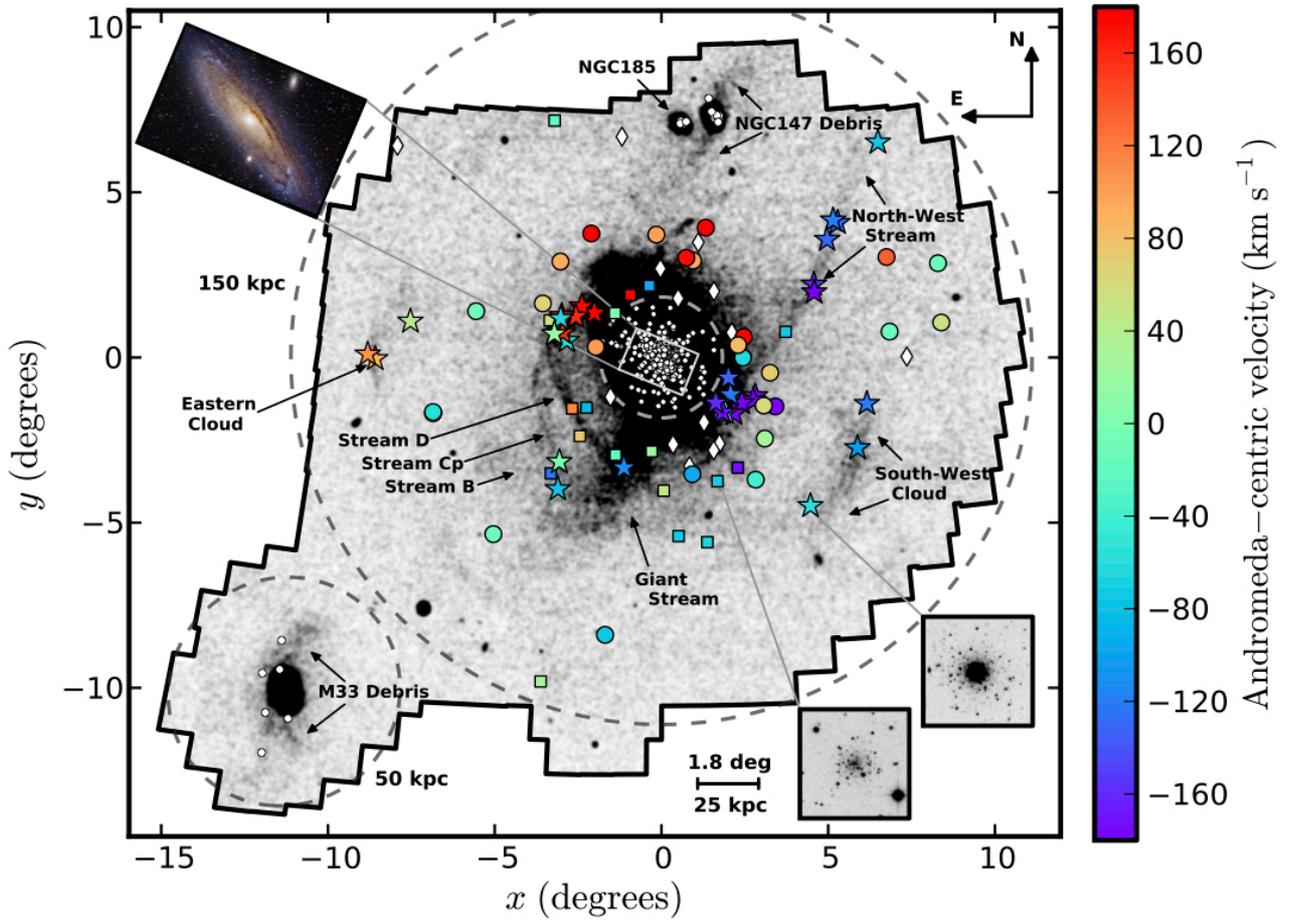

**Figure 1.** Map showing the distribution of metal-poor red giant stars in Andromeda's halo. The density has been statistically corrected for contamination, and smoothed with a Gaussian kernel. Members of Andromeda's globular cluster system are marked with filled points. Those outside 25 kpc are coloured according to their line-of-sight velocities (where available) and have point styles defined by their classification (GC-Non = circles; GC-Sub = stars; ambiguous = squares; no data = diamonds). Dashed circles represent projected galactocentric radii of 25 kpc and 150 kpc, respectively. The axes define the Cartesian coordinate system used in our models; following astronomical convention, East is to the left and North is up.

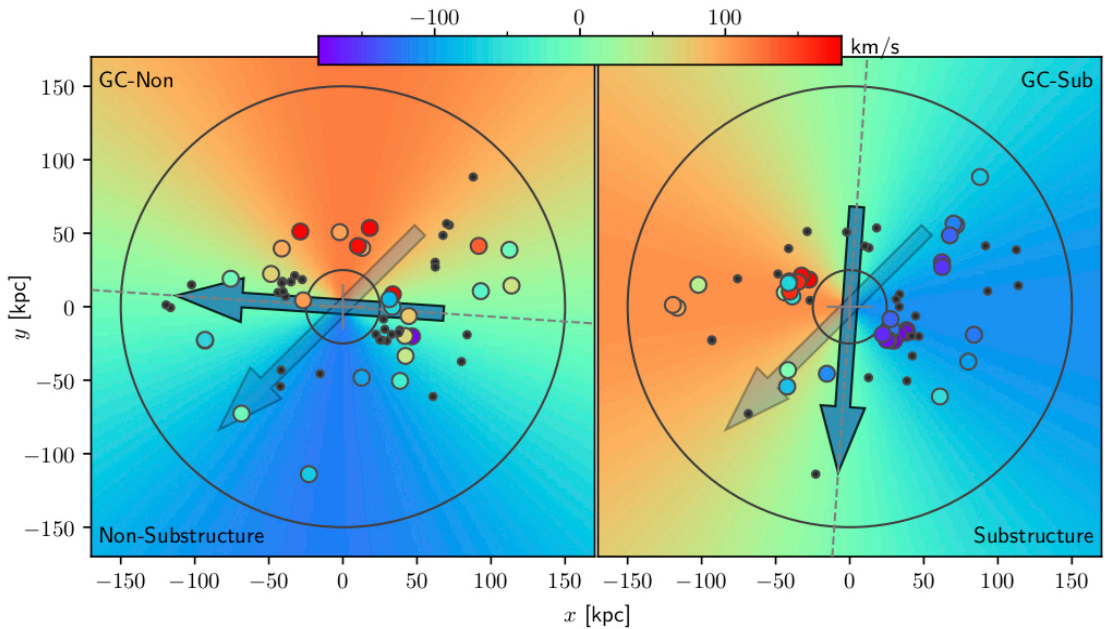

**Figure 2.** Best-fit rotational kinematics for the favoured model $V_2$. The best-fit parameters are listed in Extended Data Tables 1 and 4; the left panel shows results for GC-Non, while the right panel shows results for GC-Sub. The underlying colour illustrates the rotational kinematic signatures, while globular cluster points are coloured according to their measured M31-centric velocity. Clusters not in the applicable subsample are black points, while ambiguous clusters are not plotted but are apparent in Figure 1. Dashed lines indicate the inferred rotational axes, while arrows represent the normalized angular momentum vectors. The fainter underlying arrows show the previously-measured[7] rotation axis for the entire outer cluster sample (matching model $V_1$), which is closely aligned with that for Andromeda's disk.

# References

**1.** Searle, L. & Zinn, R. Compositions of halo clusters and the formation of the galactic halo. *The Astrophysical Journal* **225**, 357–379 (1978).

**2.** Springel, V. *et al.* Simulations of the formation, evolution and clustering of galaxies and quasars. *Nature* **435**, 629–636 (2005).

**3.** Bullock, J. S. & Johnston, K. V. Tracing Galaxy Formation with Stellar Halos. I. Methods. *The Astrophysical Journal* **635**, 931–949 (2005).

**4.** Cooper, A. P. *et al.* Galactic stellar haloes in the CDM model. *Monthly Notices of the Royal Astronomical Society* **406**, 744–766 (2010).

**5.** Johnston, K. V. *et al.* Tracing Galaxy Formation with Stellar Halos. II. Relating Substructure in Phase and Abundance Space to Accretion Histories. *The Astrophysical Journal* **689**, 936–957 (2008).

**6.** Veljanoski, J. *et al.* Kinematics of Outer Halo Globular Clusters in M31. *The Astrophysical Journal Letters* **768**, L33 (2013).

**7.** Veljanoski, J. *et al.* The outer halo globular cluster system of M31 - II. Kinematics. *Monthly Notices of the Royal Astronomical Society* **442**, 2929–2950 (2014).

**8.** Perrett, K. M. *et al.* The Kinematics and Metallicity of the M31 Globular Cluster System. *The Astronomical Journal* **123**, 2490–2510 (2002).

**9.** Caldwell, N. & Romanowsky, A. J. Star Clusters in M31. VII. Global Kinematics and Metallicity Subpopulations of the Globular Clusters. *The Astrophysical Journal* **824**, 42 (2016).

**10.** Ferguson, A. M. N., Irwin, M. J., Ibata, R. A., Lewis, G. F. & Tanvir, N. R. Evidence for Stellar Substructure in the Halo and Outer Disk of M31. *The Astronomical Journal* **124**, 1452–1463 (2002).

**11.** Ibata, R. A. *et al.* The Haunted Halos of Andromeda and Triangulum: A Panorama of Galaxy Formation in Action. *The Astrophysical Journal* **671**, 1591–1623 (2007).

**12.** Ibata, R. A. *et al.* The Large-scale Structure of the Halo of the Andromeda Galaxy. I. Global Stellar Density, Morphology and Metallicity Properties. *The Astrophysical Journal* **780**, 128 (2014).

**13.** Mackey, A. D. *et al.* The outer halo globular cluster system of M31 - III. Relationship to the stellar halo. *Monthly Notices of the Royal Astronomical Society* **484**, 1756–1789 (2019).

**14.** Ibata, R. A. *et al.* A vast, thin plane of corotating dwarf galaxies orbiting the Andromeda galaxy. *Nature* **493**, 62–65 (2013).

**15.** McConnachie, A. W. *et al.* The remnants of galaxy formation from a panoramic survey of the region around M31. *Nature* **461**, 66–69 (2009).

**16.** McConnachie, A. W. *et al.* The Large-scale Structure of the Halo of the Andromeda Galaxy. II. Hierarchical Structure in the Pan-Andromeda Archaeological Survey. *The Astrophysical Journal* **868**, 55 (2018).

**17.** Huxor, A. P. *et al.* Globular clusters in the outer halo of M31: the survey. *Monthly Notices of the Royal Astronomical Society* **385**, 1989–1997 (2008).

**18.** Huxor, A. P. *et al.* The outer halo globular cluster system of M31 - I. The final PAndAS catalogue. *Monthly Notices of the Royal Astronomical Society* **442**, 2165–2187 (2014).

**19.** Mackey, A. D. *et al.* Evidence for an Accretion Origin for the Outer Halo Globular Cluster System of M31. *The Astrophysical Journal Letters* **717**, L11–L16 (2010).

**20.** Ibata, R. A. *et al.* On the Accretion Origin of a Vast Extended Stellar Disk around the Andromeda Galaxy. *The Astrophysical Journal* **634**, 287–313 (2005).

**21.** Dorman, C. E. *et al.* A Clear Age-Velocity Dispersion Correlation in Andromeda's Stellar Disk. *The Astrophysical Journal* **803**, 24 (2015).

**22.** Ibata, R. A., Irwin, M., Lewis, G., Ferguson, A. M. N. & Tanvir, N. A giant stream of metal-rich stars in the halo of the galaxy M31. *Nature* **412**, 49–52 (2001).

**23.** Fardal, M. A. *et al.* Inferring the Andromeda Galaxy's mass from its giant southern stream with Bayesian simulation sampling. *Monthly Notices of the Royal Astronomical Society* **434**, 2779–2802 (2013).

**24.** Hammer, F. *et al.* A 2–3 billion year old major merger paradigm for the Andromeda galaxy and its outskirts. *Monthly Notices of the Royal Astronomical Society* **475**, 2754–2767 (2018).

**25.** D'Souza, R. & Bell, E. F. The Andromeda galaxy's most important merger about 2 billion years ago as M32's likely progenitor. *Nature Astronomy* **2**, 737–743 (2018).

**26.** Conn, A. R. *et al.* The Three-dimensional Structure of the M31 Satellite System; Strong Evidence for an Inhomogeneous Distribution of Satellites. *The Astrophysical Journal* **766**, 120 (2013).

**27.** Fernando, N. *et al.* On the stability of satellite planes - I. Effects of mass, velocity, halo shape and alignment. *Monthly Notices of the Royal Astronomical Society* **465**, 641–652 (2017).

**28.** Fernando, N., Arias, V., Lewis, G. F., Ibata, R. A. & Power, C. Stability of satellite planes in M31 II: effects of the dark subhalo population. *Monthly Notices of the Royal Astronomical Society* **473**, 2212–2221 (2018).

**29.** Libeskind, N. I. *et al.* Planes of satellite galaxies and the cosmic web. *Monthly Notices of the Royal Astronomical Society* **452**, 1052–1059 (2015).

**30.** Harris, W. E., Blakeslee, J. P. & Harris, G. L. H. Galactic Dark Matter Halos and Globular Cluster Populations. III. Extension to Extreme Environments. *The Astrophysical Journal* **836**, 67 (2017).

**31.** Martin, N. F. *et al.* The PAndAS View of the Andromeda Satellite System. I. A Bayesian Search for Dwarf Galaxies Using Spatial and Color-Magnitude Information. *The Astrophysical Journal* **776**, 80 (2013).

**32.** Galleti, S., Federici, L., Bellazzini, M., Fusi Pecci, F. & Macrina, S. 2MASS NIR photometry for 693 candidate globular clusters in M 31 and the Revised Bologna Catalogue. *Astronomy & Astrophysics* **416**, 917–924 (2004).

**33.** Caldwell, N. *et al.* Star Clusters in M31. I. A Catalog and a Study of the Young Clusters. *The Astronomical Journal* **137**, 94–110 (2009).

**34.** Peacock, M. B. *et al.* The M31 globular cluster system: ugriz and K-band photometry and structural parameters. *Monthly Notices of the Royal Astronomical Society* **402**, 803–818 (2010).

**35.** Caldwell, N., Schiavon, R., Morrison, H., Rose, J. A. & Harding, P. Star Clusters in M31. II. Old Cluster Metallicities and Ages from Hectospec Data. *The Astronomical Journal* **141**, 61 (2011).

**36.** Dorman, C. E. *et al.* A New Approach to Detailed Structural Decomposition from the SPLASH and PHAT Surveys: Kicked-up Disk Stars in the Andromeda Galaxy? *The Astrophysical Journal* **779**, 103 (2013).

**37.** Brewer, B. J., Pártay, L. B. & Csányi, G. Diffusive nested sampling. *Statistics and Computing* **21**, 649–656 (2011).

**38.** Brewer, B. J. & Foreman-Mackey, D. DNest4: Diffusive nested sampling in C++ and python. *Journal of Statistical Software, Articles* **86**, 1–33 (2018).

**39.** Skilling, J. Nested sampling for general Bayesian computation. *Bayesian Analysis* **1**, 833–859 (2006).

**40.** Foreman-Mackey, D. corner.py: Scatterplot matrices in python. *The Journal of Open Source Software* **1**, 24 (2016).

## Acknowledgements

This work is based in part on observations obtained with MegaPrime/MegaCam, a joint project of CFHT and CEA/DAPNIA, at the Canada-France-Hawaii Telescope (CFHT) which is operated by the National Research Council (NRC) of Canada, the Institut National des Sciences de l'Univers of the Centre National de la Recherche Scientifique of France, and the University of Hawaii.

This work is further based in part on observations obtained at the Gemini Observatory, which is operated by the Association of Universities for Research in Astronomy, Inc., under a cooperative agreement with the NSF on behalf of the Gemini partnership: the National Science Foundation (United States), National Research Council (Canada), CONICYT (Chile), Ministerio de Ciencia, Tecnología e Innovación Productiva (Argentina), Ministério da Ciência, Tecnologia e Inovação (Brazil), and Korea Astronomy and Space Science Institute (Republic of Korea).

Some of the data presented herein were obtained at the W. M. Keck Observatory, which is operated as a scientific partnership among the California Institute of Technology, the University of California and the National

Aeronautics and Space Administration. The Observatory was made possible by the generous financial support of the W. M. Keck Foundation.

The authors wish to recognise and acknowledge the very significant cultural role and reverence that the summit of Maunakea has always had within the indigenous Hawaiian community. We are most fortunate to have the opportunity to conduct observations from this mountain.

The William Herschel Telescope (WHT) is operated on the island of La Palma by the Isaac Newton Group of Telescopes in the Spanish Observatorio del Roque de los Muchachos of the Instituto de Astrofísica de Canarias.

This work is based in part on observations at Kitt Peak National Observatory, National Optical Astronomy Observatory, which is operated by the Association of Universities for Research in Astronomy (AURA) under cooperative agreement with the National Science Foundation. The authors are honoured to be permitted to conduct astronomical research on Iolkam Du'ag (Kitt Peak), a mountain with particular significance to the Tohono O'odham.

DM is supported by an Australian Research Council (ARC) Future Fellowship (FT160100206). GFL acknowledges support from a Partnership Collaboration Award between the University of Sydney and the University of Edinburgh. DM and GFL appreciate the hospitality of the Royal Observatory, Edinburgh where the final stages of the preparation of this paper were undertaken. BJB thanks the Marsden Fund of the Royal Society of New Zealand. This work has been published under the framework of the IdEx Unistra and benefits from a funding from the state managed by the French National Research Agency as part of the investments for the future program. ZW is supported by a Dean's International Postgraduate Research Scholarship at the University of Sydney.

## Author Contributions

AWM, RAI, MJI, AMNF, GFL and NT initiated the Pan-Andromeda Archaeological Survey (PAndAS), with extensive data analysis and interpretation undertaken with NM, MLMC, PC and JP. DM, AMNF, JV, and APH were responsible for the discovery and characterisation of PAndAS globular clusters, for measuring their line-of-sight velocities, and for conducting detailed earlier analyses of this population. GFL was responsible for the development of the kinematic models employed in this study. GFL and BJB undertook the statistical analysis, with BJB responsible for implementing and running the kinematic models in DNest4. ZW was responsible for undertaking geometric transformations into the Andromeda frame to enable comparisons with previous studies. All authors assisted in the interpretation of the results and writing of the paper.

## Additional Information

Reprints and permissions information is available at www.nature.com/reprints . The authors declare no competing interests. Correspondence and requests for materials should be addressed to DM (email: dougal.mackey@anu.edu.au ).

## Methods

**Observational data.** The globular clusters used in this study were discovered as part of the Pan-Andromeda Archaeological Survey (PAndAS), a multi-band panoramic imaging program targeting the Andromeda (M31) and Triangulum (M33) galaxies out to physical distances (in projection) of ∼ 150 kpc and ∼ 50 kpc respectively, which used the MegaCam wide-field camera on the 3.6m Canada-France-Hawaii Telescope[15,16,31]. We focus on the 92 objects known at projected galactocentric radii larger than 25 kpc[13]. Our motivation for selecting this as the inner radial limit for our analysis is discussed in more detail below.

Line-of-sight velocities for globular clusters in this sample were obtained through a series of spectroscopic observations, using ISIS on the 4.2m William Herschel Telescope at the Observatorio del Roque de los Muchachos on the island of La Palma, RC at the 4m Mayall Telescope at the Kitt Peak National Observatory in Arizona, GMOS on the 8m Gemini North Telescope in Hawaii, and DEIMOS on the 10m Keck II Telescope, also in Hawaii. The full set of outer halo clusters with measured velocities numbers 77 objects of the 92 known outside 25 kpc[6,7,13]. For the present analysis, the line-of-sight velocities were transformed into the Galactocentric frame and then an M31-centred frame using the procedure outlined in these previous contributions. This accounts for projection effects due to both the Solar motion and the systemic velocity of M31, induced by the large angular span of the M31 halo on the sky. The typical measurement uncertainty per cluster is ∼ 10 km/s.

Overall, some ∼ 500 globular clusters have been catalogued in M31[9,13,32–35], the majority of which (≈ 80%) sit at radii smaller than our inner limit of 25 kpc. Another ∼ 50 clusters are known to be members of large Andromedan satellites such as M33, NGC 147, NGC 185, and NGC 205[13]. While these various cluster populations were excluded from our analysis, we utilised these previous compilations to plot them in Figure 1 for context.

**Relationship to stellar substructure.** Details of the relationship between the outer globular cluster population and the underlying stellar halo in Andromeda were developed in a recent publication[13]. In summary, the clusters can be assigned to one of three groups: those that exhibit strong spatial and/or kinematic evidence for a link with a substructure in the field halo, those that show no such evidence, and those for which the evidence is weak or conflicting. Throughout this paper, these three sets are labelled as "substructure globular clusters" (GC-Sub), "non-substructure globular clusters" (GC-Non), and "ambiguous globular clusters", respectively.

"Spatial" evidence is quantified by measuring the density of halo stars locally around each cluster, and finding those objects with high local densities relative to the observed distribution at commensurate galactocentric radii. "Kinematic" evidence is limited to identifying objects that are members of small-scale dynamically cold groupings; no information about global velocity patterns is utilised. For the 77 clusters considered here, this classification scheme yields subgroups of 32 objects in GC-Sub and 26 in GC-Non, with the remaining 19 identified as ambiguous. In a later section we detail how these ambiguous globular clusters are incorporated into our models.

The inner halo of M31 is highly complex. At radii smaller than 25 kpc, the stellar substructures become so pervasive, and the degree of overlap so great, that the classification scheme outlined above breaks down[13]. Moreover, the dynamical times become sufficiently short that the association between stellar substructures and globular clusters from accreted satellites is far less persistent than at larger radii. This inner region is also complicated by the presence of extensive stellar material disrupted from the disk of Andromeda[20,21,36], as well as a substantial *in situ* globular cluster population (i.e., non-accreted objects that formed in Andromeda's disk or bulge)[9]. These issues motivate our choice of 25 kpc as the inner radius defining the sample for analysis; our key results are not sensitive to mild variations in the adopted limit (∼ 20 – 40 kpc).

**Kinematic models.** To explore the velocity characteristics of M31's outer globular clusters, we consider models in which the population possesses an overall velocity dispersion superimposed upon an underlying net rotation. We employ three physically-motivated models for the rotational component, which are expressed in detail in Extended Data Table 1. The $V$ models, which were used in a previous study on the bulk rotation of globular cluster systems[6,7], consider a constant amplitude of rotational velocity modulated by an angular dependence. The $S$ models represent solid-body rotation, where the rotational velocity increases linearly from the axis of rotation. Finally, the $F$ models consider an asymptotically-flat velocity profile with a smooth transition through the axis of rotation, akin to the rotation curves of spiral galaxies. These models represent three broad families of potential rotation and, as will be seen in detail below, the choice of a specific model does not impact the need for two rotational components to explain the observed velocity properties of Andromeda's outer globular cluster population.

For each rotational model, we consider two distinct scenarios. Initially, we treat the entire globular cluster population as a whole, considering a single underlying global rotation[6,7]. Following this, we treat GC-Sub and GC-Non as distinct populations, each possessing their own characteristic rotational component.

In addition to the rotational component, the models also include a velocity dispersion as a function of position given by

$$\Sigma(r, \theta; \Lambda) = \sigma \left(\frac{r}{R_0}\right)^\gamma$$

where $r$ is the projected galactocentric radius, $\theta$ is the angular coordinate in the chosen system (detailed below), $\gamma$ is a slope parameter and $\sigma$ is the velocity dispersion at projected radius $R_0$; to compare with previous work, and as this parameter is degenerate with $\sigma$ and $\gamma$, we fix $R_0 = 30$ kpc[6,7]. The model parameters to be estimated, including those for the rotational components of the model, are denoted collectively by $\Lambda$. This functional form for $\Sigma()$ was used for all of the rotational models. For the models where we consider two distinct rotational components, we also consider each to have their own distinct dispersion profiles as detailed above.

**Bayesian inference.** To test these models for the rotational pattern exhibited by the globular clusters, we employ Bayesian inference to compute the probabilities of the models given the data. Clearly, the models have differing numbers of parameters, but we can employ the marginal likelihood (also known as the evidence) to test the various propositions.

Throughout this paper we adopt a Cartesian coordinate system in the tangent plane centred on M31, with the x-axis aligned with West and the y-axis with North in the equatorial astronomical system (see Figure 1). In detailing our approach, we let ($r_i$, $\theta_i$) be the position of the $i^{th}$ globular cluster on the sky, in plane polar coordinates with respect to the above coordinate system, and let $v_i$ be its measured velocity along the line of sight, including rotation, velocity dispersion, and measurement error. Our model attempts to predict $v_i$ as a function of ($r_i$, $\theta_i$), in a manner analogous to regression. The rotational component is specified in terms of a functional form, $f(r, \theta; \Lambda)$, which gives the model-predicted velocity as a function of position. This is combined with the model-predicted line-of-sight velocity dispersion, $\Sigma(r, \theta; \Lambda)$, to specify the full kinematic model.

The probability distribution for the data (i.e., the *measured* velocities) $v_i$, given the parameters and one of the models from Extended Data Table 1, is

$$v_i \mid \Lambda \sim \text{Normal}(f(r, \theta; \Lambda), \Sigma(r, \theta; \Lambda)^2 + s_i^2).$$

where the $\{s_i\}$ quantities are the uncertainties on the velocity measurements $\{v_i\}$. With this choice, the likelihood function can be written explicitly as

$$L(\Lambda) = \prod_{i=1}^{N} \frac{1}{\sqrt{2\pi(\Sigma(r_i, \theta_i; \Lambda)^2 + s_i^2)}} \exp\left(-\frac{(v_i - f(r_i, \theta_i; \Lambda))^2}{2(\Sigma(r_i, \theta_i; \Lambda)^2 + s_i^2)}\right)$$

where $N$ is the total number of globular clusters in the sample. In this paper, *parameter estimation* involves computing the posterior distributions for the parameters $\Lambda$ given the data assuming a model from Extended Data Table 1 (i.e., given a particular choice for $f()$), and *model comparison* involves the assessment of different choices for the form of the function $f()$.

**Handling of ambiguous globular clusters.** For 19 out of the 77 clusters, it is unclear whether or not a substructure is present. For each of these ambiguous objects, we assign a flag $f_j$ which takes the value of 1 if the cluster really is associated with a substructure and 0 if it is not. We set the prior for the $\{f_j\}$ parameters by introducing a hyperparameter $p_{\text{subs}}$, which sets the prior expected proportion of the ambiguous globular clusters that actually are associated with a substructure. The prior for $p_{\text{subs}}$ and the $\{f_j\}$ flags is then

$$p_{\text{subs}} \sim \text{Uniform}(0,1)$$

$$f_j \sim \text{Bernoulli}(p_{\text{subs}}).$$

We have also tested alternative approaches for treating the ambiguous globular clusters, such as omitting them completely from the sample, and found this has only minor effects on the results. Hence our conclusions are not driven by the treatment of the ambiguous clusters.

**Prior distributions.** The joint prior distribution for all parameter and data values reflects all the assumed information used in the analysis apart from the data itself. This is specified in Extended Data Table 2. Single-component models may be obtained by ignoring all parameters referring to the substructure component. For positive parameters of unknown order of magnitude, we have assigned truncated Cauchy priors to the logarithm of the parameter. The truncation is for numerical convenience, and the Cauchy distribution weakly emphasizes values within a few orders of unity. For two-component models, the prior for the second component implies a high probability that it is within an order of magnitude of the first component. This avoids any fatal problems caused by the 'Jeffreys-Lindley paradox' when using naïve wide priors for additional parameters in Bayesian model comparison problems.

**Computation and results.** We used DNest4[37,38] to sample the posterior distributions for the parameters of each model, and to compute the marginal likelihood of each model. The marginal likelihood of a model $M$ is the average value of the likelihood function with respect to the prior distribution for the parameters, and plays the role of a likelihood when comparing models.

The estimated marginal likelihoods are given in Extended Data Table 3. The data are far more probable under models $V_2$ and $F_2$ than the others. Within each class of models ($S$, $V$, $F$), the two-component model makes the data far more likely than the one component model. By Bayes's theorem, the two-component models are thus much more probable than the one-component models assuming they have nontrivial prior probabilities. If prior probabilities of 1/6 are assigned to each model, the posterior probability for two substructures (i.e., for the proposition $V_2 \vee S_2 \vee F_2$) is 0.9994. Crucially, this is conditional on $\top \equiv S_1 \vee S_2 \vee V_1 \vee V_2 \vee F_1 \vee F_2$ (i.e., the proposition that *it is one of these models*), and other sets of assumptions may always be considered. The marginal likelihood of $\top$ is −468.51, which can be used to compare our findings with any future analysis[39].

Estimates for the parameters in the overall favoured model, $V_2$, are reported in Extended Data Table 4 and were used in the construction of Figure 2. A corner plot showing the marginal distributions for the parameters is provided in Extended Data Figure 1. We note that the second-best model by marginal likelihood, $F_2$, whilst still favouring a dual rotational component model over a single rotational component, possesses a slightly different angle for the orientation of GC-Sub, that is, $\phi_1 \sim 60°$. Under $F_2$, the orientation of GC-Non is $\phi_0 = 0°$, consistent with the other models. Investigating the posterior distributions derived for $F_2$ reveals that the preferred scale parameter, $L$, becomes unphysically small, resulting in a step function in velocity across the rotational axis. This results in step functions in the likelihood, with the few individual globular clusters close to the rotation axis strongly constraining $\phi_1$.

The analysis in this paper considered the 77 globular clusters with measured velocities, from an overall sample of 92. We note, therefore, that the sample is incomplete. However, examining the marginal distributions presented in Extended Data Figure 1, we see that that the velocity amplitudes are strongly constrained to be non-zero, and similarly the orientations are precisely constrained. It is unlikely that the velocities for clusters in the remaining sample are extreme enough to counter these outcomes.

# Data Availability

All data analysed for this study are publicly available. PAndAS data products, including the stellar photometry catalogue, reduced individual images, and image stacks, may be downloaded from the Canadian Astronomical Data Center (CADC)[16]. Globular cluster locations, radial velocities, and classifications are published online[7,13,18] and are also found in the code repository, linked below.

# Code Availability

Readers may access the code used for the inference, with the data included, at https://github.com/eggplantbren/AndromedaMixture . The code has been released under the permissive MIT license. The README describes how to reproduce the results of the paper.

# Extended Data

| Model | Description | Functional form of $f()$ |
|---|---|---|
| $V_1$ | Veljanoski et al., one component | $A_0 \sin(\theta - \phi_0)$ |
| $V_2$ | Veljanoski et al., two components | $A_k \sin(\theta - \phi_k)$ |
| $S_1$ | Solid-body, one component | $A_0 \left( x \sin \phi_0 - y \cos \phi_0 \right)$ |
| $S_2$ | Solid-body, two components | $A_k \left( x \sin \phi_k - y \cos \phi_k \right)$ |
| $F_1$ | Asymptotically-flat rotation curve | $A_0 \tanh\left( (x \sin \phi_0 - y \cos \phi_0)/L_0 \right)$ |
| $F_2$ | Asymptotically-flat, two components | $A_k \tanh\left( (x \sin \phi_k - y \cos \phi_k)/L_k \right)$ |
| $\top$ | Disjunction of all of the above | |

**Extended Data Table 1.** Functional form of the rotational component for the models under consideration. The third column gives the model-predicted line of sight velocity as a function of position on the sky, expressed in a combination of Cartesian coordinates (*x*, *y*) and plane polar coordinates (*r*, *θ*) defined as specified in the text. For two-component models the subscript *k* takes values of 0 for GC-Non and 1 for GC-Sub. For one-component models, only subscript 0 applies.

| Quantity | Meaning | Distribution |
| --- | --- | --- |
| **Parameters** | | |
| $A_0$ | Amplitude of rotation | $\ln(A_0) \sim \text{Cauchy}(0, 5)T(-100, 100)$ |
| $A_1$ | Amplitude of rotation | $\ln(A_1) \sim \text{Normal}(\ln A_0, 1)$ |
| $\phi_0$ | Orientation of rotation axis | $\text{Uniform}(-180°, 180°)$ |
| $\phi_1$ | Orientation of rotation axis | $\text{Uniform}(-180°, 180°)$ |
| $\sigma_0$ | Velocity dispersion at distance $R_0$ | $\ln \sigma_0 \sim \text{Cauchy}(0, 5)T(-100, 100)$ |
| $\sigma_1$ | Velocity dispersion at distance $R_0$ | $\ln \sigma_0 \sim \text{Normal}(\ln \sigma_0, 1)$ |
| $\gamma_0$ | Velocity dispersion slope | $\text{Uniform}(-3, 0)$ |
| $\gamma_1$ | Velocity dispersion slope | $\text{Uniform}(-3, 0)$ |
| $p_{\text{subs}}$ | Hyperparameter for $\{f_j\}$ | $\text{Uniform}(0, 1)$ |
| $\{f_j\}$ | Substruct. flags for ambiguous GCs | $\text{Bernoulli}(p_{\text{subs}})$ |
| $L_0$ | Length scale of transition | $\ln L_0 \sim \text{Uniform}(\ln(10^{-3} R_0), \ln R_0)$ |
| $L_1$ | Length scale of transition | $\ln L_1 \sim \text{Uniform}(\ln(10^{-3} R_1), \ln R_1)$ |
| **Data** | | |
| $\{v_i\}$ | Measured line-of-sight GC velocities | Equation 2 |
| **Prior info.** | | |
| $N$ | Number of GCs | Given (77) |
| $\{(r_i, \theta_i)\}$ | Positions of GCs on sky | Given |
| $\{g_i\}$ | Classification flags for all GCs | Given (GC-Non / GC-Sub / Ambiguous) |

**Extended Data Table 2.** The joint prior distribution for all parameters and data. For two-component models, subscript 0 corresponds to GC-Non and subscript 1 applies to GC-Sub. For one-component models, only subscript 0 applies. The length scale parameters $L_0$ and $L_1$ only apply to models $F_1$ and $F_2$ and have no effect in the other models. The orientation angles $\phi$ are defined according to the Cartesian coordinate system specified in the text, which differs from the astronomical convention of angles measured North through East by 90°.

| Model | Seed | $\ln(Z)$ | $Z/Z_{\max}$ |
|---|---|---|---|
| $V_1$ | 0 | $-474.41$ | $7.85 \times 10^{-4}$ |
| $V_2$ | 1 | $-467.26$ | 1 |
| $S_1$ | 2 | $-479.64$ | $4.20 \times 10^{-6}$ |
| $S_2$ | 3 | $-475.29$ | $3.26 \times 10^{-4}$ |
| $F_1$ | 4 | $-475.45$ | $2.77 \times 10^{-4}$ |
| $F_2$ | 5 | $-467.59$ | 0.719 |
| $\top$ | | $-468.51$ | |

**Extended Data Table 3.** Marginal likelihoods for each of the models, along with the resulting Bayes Factor compared to the most favoured model $V_2$. Models $V_2$ and $F_2$ are strongly favoured by the data. "Seed" refers to the random number generator seed employed in each run of the MCMC exploration, and is provided for reproducibility reasons.

| Parameter | Post. Mean ± Post. SD | Max. Likelihood |
|---|---|---|
| $A_0$ (km/s) | $92 \pm 27$ | 120 |
| $A_1$ (km/s) | $110 \pm 15$ | 110 |
| $\phi_0$ (°) | $-6.2 \pm 13.4$ | $-3.9$ |
| $\phi_1$ (°) | $80.0 \pm 10.0$ | 86.0 |
| $\sigma_0$ (km/s) | $134 \pm 36$ | 101 |
| $\sigma_1$ (km/s) | $133 \pm 30$ | 130 |
| $\gamma_0$ | $-0.66 \pm 0.32$ | $-0.52$ |
| $\gamma_1$ | $-0.84 \pm 0.33$ | $-0.95$ |
| $p_{\text{subs}}$ | $0.67 \pm 0.20$ | 0.60 |

**Extended Data Table 4.** Estimates of the parameters for the favoured model $V_2$. These are reported as the posterior mean ± the posterior standard deviation. Since none of the marginal posterior distributions are extremely asymmetric or non-gaussian (Extended Data Figure 1), these are approximately 68% credible intervals. As noted previously, a subscript of 0 corresponds to GC-Non and 1 to GC-Sub. The maximum likelihood estimates, given in the final column, were used in the construction of Figure 2.

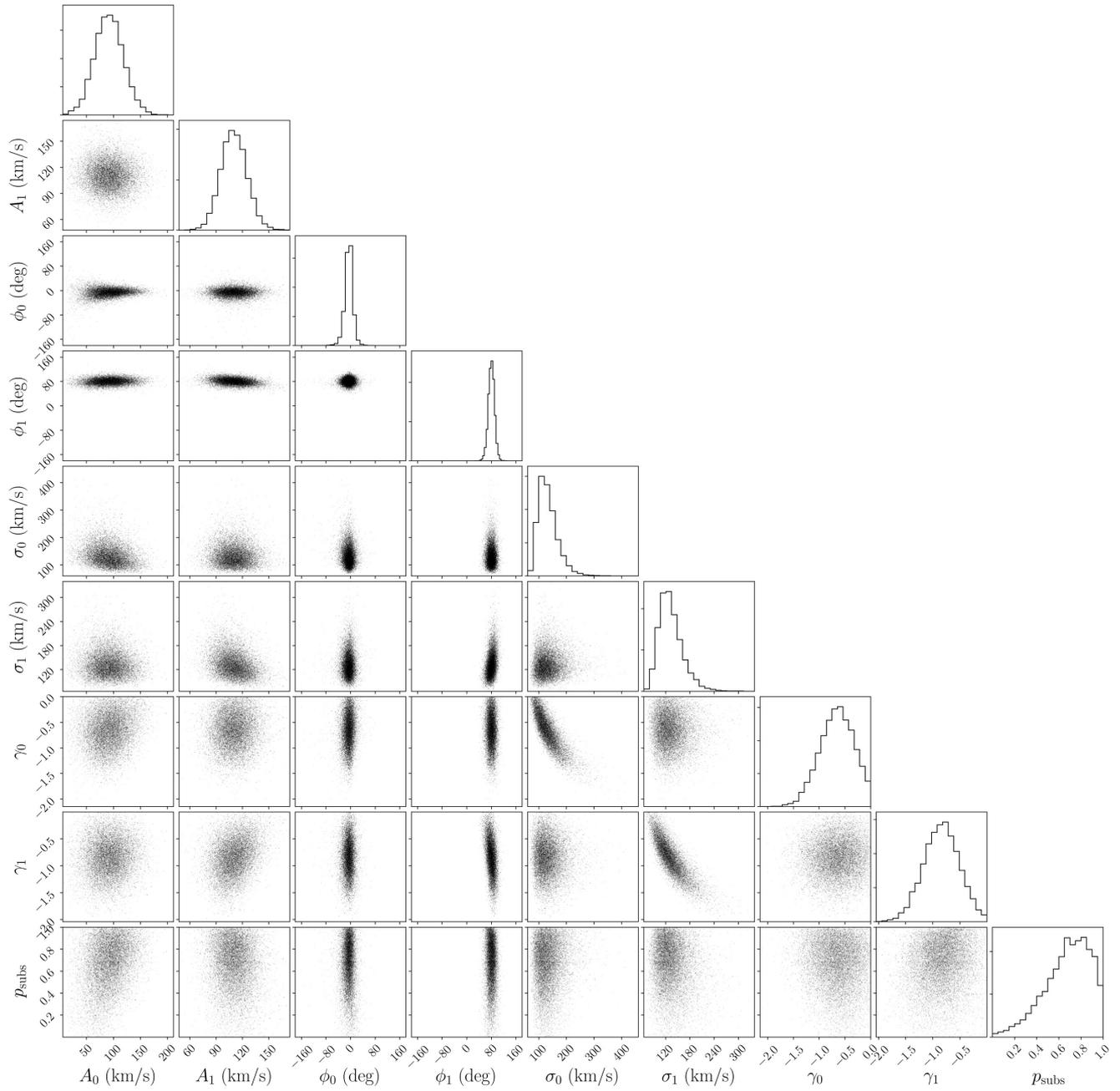

**Extended Data Figure 1.** Corner plot[40] of the posterior distribution of the parameters for the $V_2$ model. This model is defined in Extended Data Table 1. Subscript 0 and 1 correspond to the parameters for the non-substructure (GC-Non) and substructure (GC-Sub) samples respectively, and a summary of the marginal distributions is given in Extended Data Table 4.